# Rethinking Cryptophane-A for Methane Gas Sensing: Cross-Sensitivity to N₂ and CO₂ at Ambient Conditions

*Sebastián Alberti*[1], Thierry Brotin[2], Jana Jágerská[1]*

[1]Department of Physics and Technology, UiT The Arctic University of Norway, NO-9037, Tromsø, Norway.

[2]ENS de Lyon, CNRS, UMR 5182, Laboratoire de Chimie, Univ Lyon, Lyon 69342, France

ABSTRACT. Since the affinity of Cryptophane-A for methane was first reported in 1993, cryptophane-doped polymer films have been extensively studied as enrichment cladding layers in plasmonic, fiber-optic, and integrated waveguide-based optical sensors. While the use of cryptophane-doped layers has improved methane sensitivity compared to undoped claddings, controversy has grown over the years regarding their claimed selectivity and practical applicability. Key questions remain unresolved, including the extent of true methane enrichment at room temperature, cross-sensitivity to other gases, and the proportion of active cryptophane molecules within the polymer matrix. In this work, we employ Raman spectroscopy to provide direct and unambiguous evidence that Cryptophane-A exhibits measurable affinity for major ambient gases other than methane at room temperature. Notably, both nitrogen and carbon dioxide are shown to enter the cryptophane cavity and compete with methane for binding sites. This study underscores the value of Raman spectroscopy as a benchmark technique for investigating gas capture within host molecules at ambient conditions. It offers deeper insight into the binding behavior of Cryptophane-A and enables quantification of its relative affinities to atmospheric gases, thereby revealing both the limitations and the potential of cryptophane for future sensing applications.

**Introduction**

Cryptophanes, a class of hollow organic molecules with a well-defined internal cavities, have attracted attention for their ability to trap small molecules,[1,2] ions,[3] and even weakly interacting gases molecules.[4] For instance, the complexation of methane has inspired the development of cryptophane-doped cladding materials for methane sensing. When applied to non-specific sensors such as optical fiber gratings,[5] photonic crystals,[6] Mach–Zehnder interferometers (MZI)[7], surface plasmon resonance (SPR)[8], or quartz crystal microbalances (QCM) sensor[9], these claddings have shown a significant methane enrichment, and, consequently, enhanced sensitivity compared to unclad sensors or polymer claddings without cryptophanes. Despite this progress, growing evidence from recent studies points to the intricate and sometimes problematic behaviour of cryptophanes in practical sensing applications. One key issue is the conformational change, which can reduce the amount of active cryptophane sites, or generate new moieties that need to be evaluated. These structural alterations can decrease the availability of cryptophanes for guest binding, thereby diminishing the enrichment capability of the cladding and its practical utility. Conformational changes, such as imploded cryptophanes[10], were proposed decades ago[11] and have recently been confirmed experimentally.[12] Even more concerning is the growing number of reported interferences or competing guest molecules. Substances such as halomethanes, water,[13] ethanol,[8] nitrous oxide,[9] ammonia[14] and even hydrogen[15] have all been identified as potential interferents, and new ones are continually being added. This increasing list raises questions about the assumed selectivity/specificity of cryptophane-doped claddings.

Interferences are key detrimental factors in non-specific sensing setups, such as refractive index sensors, which are used in most studies involving cryptophane-doped claddings. In these systems, changes in temperature, pressure, humidity, or general adsorption of analytes to the polymer matrix - rather than to the cryptophane itself - can produce a measurable signal (see Supplementary Information S1). As well, physical changes in the polymer, such as aging or swelling due to humidity, can further confound the measurements. While reference sensors, e.g., identical devices without cryptophanes in the cladding, can help mitigate these effects, inconsistent results from both optical refractive index sensor (eg. MZI, SPR) and other non-specific sensors such as QCM[9] or contact potential Difference (CPD)[16] suggest that cryptophane cavity itself may be subject to interference from atmospheric gasses beyond methane.

Traditionally, Proton Nuclear Magnetic Resonance (H-NMR) spectroscopy has been the benchmark technique for studying Methane-cryptophane binding. However, this approach is constrained by the requirement for low temperatures to resolve the H-NMR signal of the methane–cryptophane complex, necessitating extrapolation of the thermodynamic parameters to ambient conditions. Additionally, alternative guest molecules pose further complications. For example, $H_2$ complexes are considered non detectable in normal operation conditions, $N_2$ and $CO_2$ lack hydrogen atoms, and long $T_1$ relaxation times hinder $^{13}$C-NMR detection. As a result, interactions such as $N_2$ with cryptophane-000 have only been inferred indirectly,[17] while evidence for $CO_2$ - cryptophane-111 binding relies on subtle host signal splitting observed at cryogenic temperatures.[15]

In this work, we introduce Raman spectroscopy as a practical and complementary alternative to NMR. Unlike NMR, Raman spectroscopy does not depend on the presence of hydrogen and can identify and quantify aprotic gases such as $N_2$ and $CO_2$. We demonstrate that Raman spectra of $CH_4$, $N_2$, and $CO_2$ exhibit distinct shifts when these gases are encapsulated in cryptophane-A compared to their free molecular forms (Figure 1). This enables the study of host-guest complex formation, and evaluation of selectivity to the selected cryptophane. In our approach, Cryptophane-A crystals and a commercial Raman microscope were used at room temperature, highlighting the simplicity yet effectiveness of the method.

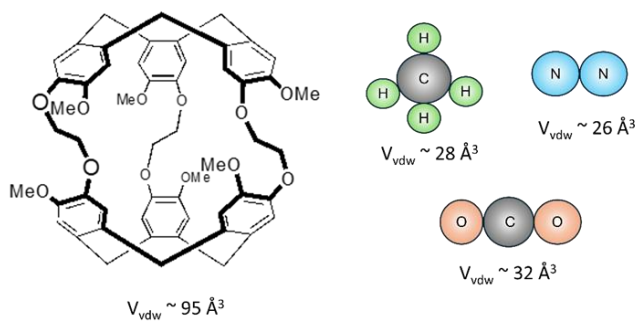

**Figure 1**. Schematic structure of Cryptophane-A, nitrogen, carbon dioxide and methane. The Van der Waals volume ($V_{vdw}$) of the cryptophane-A cavity and the respective gas molecules is also shown.

**Experimental section**

Cryptophane-A was synthesized with high purity using a multistep procedure previously reported by Brotin et al.[18] The resulting crystals were placed in a custom-built flow chamber designed to interface with a Renishaw InVia Raman microscope (Figure 2), which was then sealed using a PDMS gasket and a microscope cover glass 1.5 (160-190 um). The chamber construction facilitates the use of a 0.75 NA objective with tight focus, which provides for stronger signal from a small active volume, which can be precisely located over the cryptophane crystals thus avoiding spurious signal from the coverslip or the chamber itself. The chamber inlet and outlet from the bottom of the microscope slide allowed us to introduce the gases of interest one per time and record the Raman spectra under controlled gas atmosphere. The measurements were performed under constant flow after signal stabilization, in successive measurements with two hours in between to allow for elimination of any memory effects, such as from gases partitioned into the PDMS gasket material.

The spectrometer was configured to target specific Raman lines characteristic of the respective gases: 2917 cm$^{-1}$ for methane, 1388cm$^{-1}$ for carbon dioxide, and 2329 cm$^{-1}$ for nitrogen. These lines correspond to strong symmetric vibrational modes that produce intense Raman signals due to large changes in polarizability.[19] A 785 nm laser was used for excitation, and each spectrum was obtained by averaging 200 measurements with 2-second integration time per measurement.

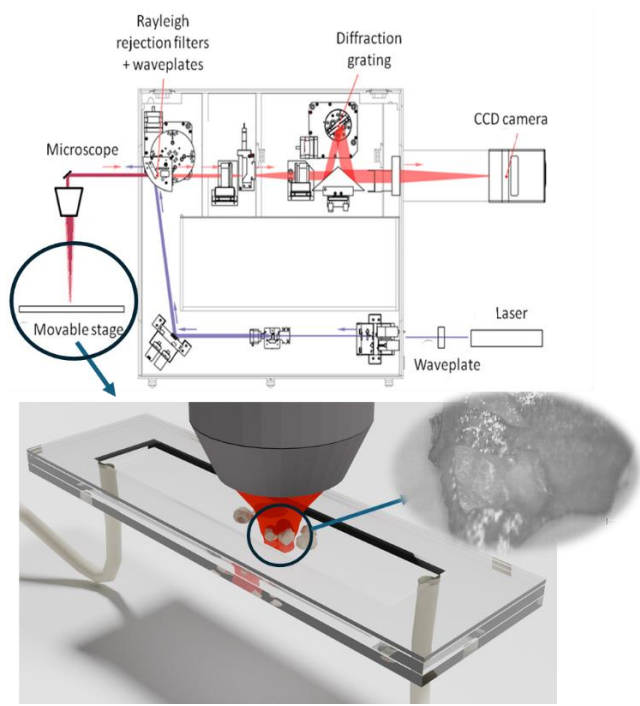

**Figure 2.** Schematic of the custom-built flow cell placed under confocal Raman microscope (Renishaw InVia). The diagram illustrates that the active volume includes both the solid crystal and the surrounding gas. Inset: image of the crystals taken with a 10X objective.

### Results and discussion

An example of Raman spectra of Cryptophane-A under $N_2$, $CO_2$, and $CH_4$ atmospheres is presented in Figure 3. For each gas, two distinct Raman peaks are observed superimposed on the background signal of Cryptophane-A. The first peak corresponds to the characteristic vibrational modes of the free gas, as reported in the literature: 2329 cm$^{-1}$ for $N_2$, 1388 cm$^{-1}$ for $CO_2$, and 2917 cm$^{-1}$ for $CH_4$. The second, noticeably red-shifted peak (i.e., at a lower wavenumber) is attributed to gas molecules encapsulated within the Cryptophane-A host structure. This red shift indicates that the trapped molecules experience a different local environment, supporting the hypothesis of host–guest interactions within the cavity.

The presence of two peaks indicates the coexistence of free and encapsulated gas within the Raman-active volume, i.e. the volume defined by the confocal volume of the objective lens equal to approximately 6.5 $\mu m^3$. Since this volume exceeds the size of the Cryptophane-A crystals, some interaction with free gas is inevitable. To enhance detection of the encapsulated species, the focus was carefully adjusted to maximize the intensity of the red-shifted peaks. Following alignment, all spectral measurements were conducted without altering the focal plane or repositioning the sample. Maintaining a consistent overlap between the free gas and solid-phase volumes was essential for comparing the relative affinities of the three gases. To improve statistical robustness, measurements were repeated at two additional locations on the sample, each characterized by slightly different gas/solid overlap conditions.

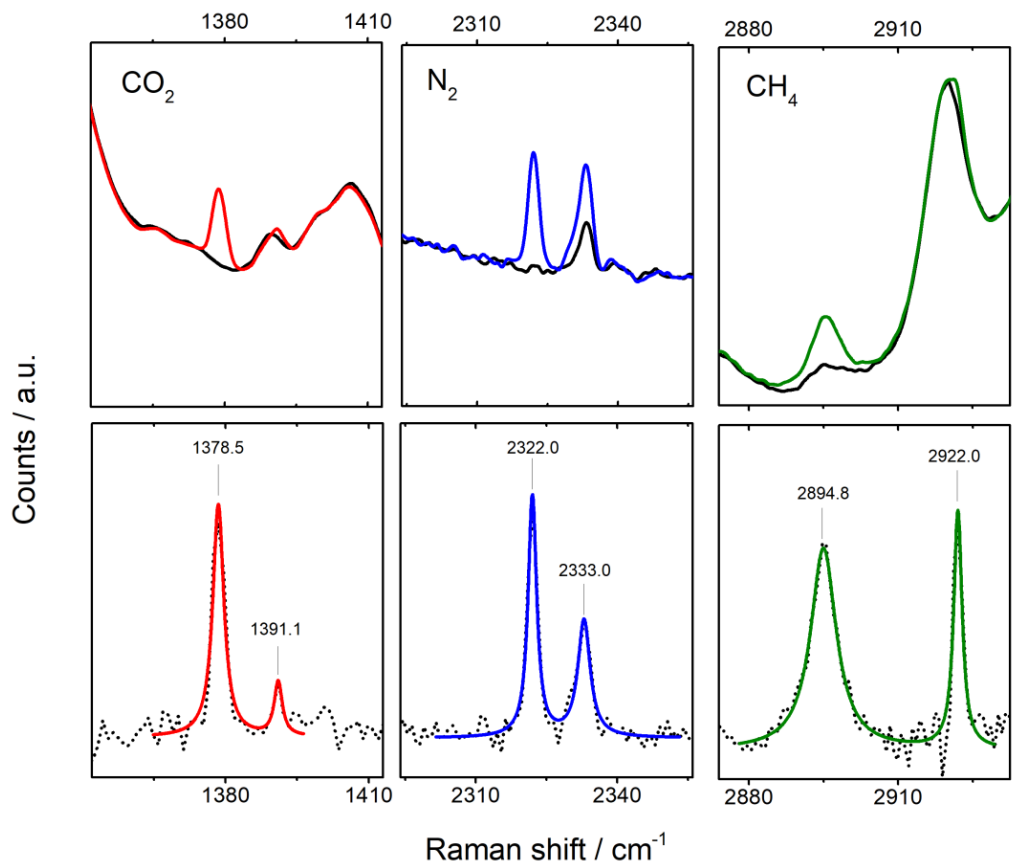

**Figure 3.** *Top row:* Raman spectra for $CO_2$, $CH_4$, and $N_2$ recorded while focusing on a Cryptophane-A crystal, before (black) and after exposure to the respective gas (color). Background spectra (black) were collected when the gas chamber was purged with an alternative gas exhibiting no features in the relevant spectral range. *Bottom row:* Background-subtracted spectra (shown as dotted lines) fitted with double Lorentzian functions (solid color lines).

For spectral analysis, background-corrected spectra were fitted using a double Lorentzian function, as shown in Figure 3. From the fit parameters, peak positions, full width at half maximum (FWHM), and red shifts are presented in Table 1 for three sample locations (denoted as (1), (2), and (3)). Integrated peak intensities are analysed in Figure 4.

**Table 1.** Raman peak positions and linewidths for three independent Cryptophane-A-($CO_2$/$CH_4$/$N_2$) measurement at different sample locations.

| Gas | Free gas position (cm-1) | Free gas width (cm-1) | Trapped gas position (cm-1) | Trapped gas width (cm-1) | Redshift (cm-1) |
|---|---|---|---|---|---|
| $CO_2$ (2) | 1391.6±0.2 | 1.5±1.0 | 1378.88±0.05 | 2.5±0.2 | 12.7±0.3 |
| CO2 (1) | 1389.8±0.2 | 1.7±0.5 | 1376.96±0.05 | 2.6±0.2 | 12.8±0.3 |
| $CO_2$ (3) | 1391.0±0.4 | 2.0±1.0 | 1378.5±0.1 | 2.8±0.3 | 12.5±0.5 |
| $CH_4$ (2) | 2922.1±0.1 | 1.8±0.3 | 2895.5±0.1 | 6.4±0.3 | 26.6±0.2 |
| CH4 (1) | 2920.9±0.1 | 1.4±0.2 | 2894.0±0.2 | 6.0±0.5 | 26.9±0.3 |
| $CH_4$ (3) | 2922.0±0.1 | 2.0±0.2 | 2895.0±0.2 | 6.3±0.5 | 27.0±0.3 |
| $N_2$ (2) | 2333.5±0.2 | 3.2±0.7 | 2322.38±0.05 | 2.4±0.2 | 11.1±0.3 |
| N2 (1) | 2331.8±0.1 | 2.8±0.3 | 2320.71±0.5 | 2.0±0.2 | 11.1±0.2 |
| $N_2$ (3) | 2332.9±0.1 | 3.1±0.4 | 2321.99±0.05 | 2.2±0.2 | 10.9±0.2 |

All three gases, $CO_2$, $CH_4$, and $N_2$, exhibit strong Raman signals in their encapsulated states, indicating that each can form a host–guest complex with Cryptophane-A. This observation suggests that competitive binding between gases is likely and challenges the notion of methane-selective sensing by Cryptophane-A-based sensors.

To estimate selectivity, the integrated intensities of the Raman peaks — i.e., areas under the peaks — were determined for both free ($A_f$) and encapsulated ($A_e$) gas. The resulting $A_e/A_f$ ratios are plotted in Figure 4 for each gas and location. While these ratios reflect the relative amounts of free and encapsulated gas within the Raman-active volume (assuming similar Raman cross sections for bound and free states), they do not directly quantify preconcentration due to uncertainties in gas–solid overlap. However, since the overlap was kept constant within each measurement location, comparing $A_e/A_f$ values between gases provides insight into the relative affinity of Cryptophane-A for each gas. This relative affinity, or selectivity, defined as the ratio between the binding constants ($K_{\text{gas 1}}/K_{\text{gas 2}}$) between gas pairs was calculated using the following expression:

$$S = \frac{K_{\text{gas 1}}}{K_{\text{gas 2}}} = \frac{\left(A_e/A_f\right)_{gas\ 1}}{\left(A_e/A_f\right)_{gas\ 2}},$$

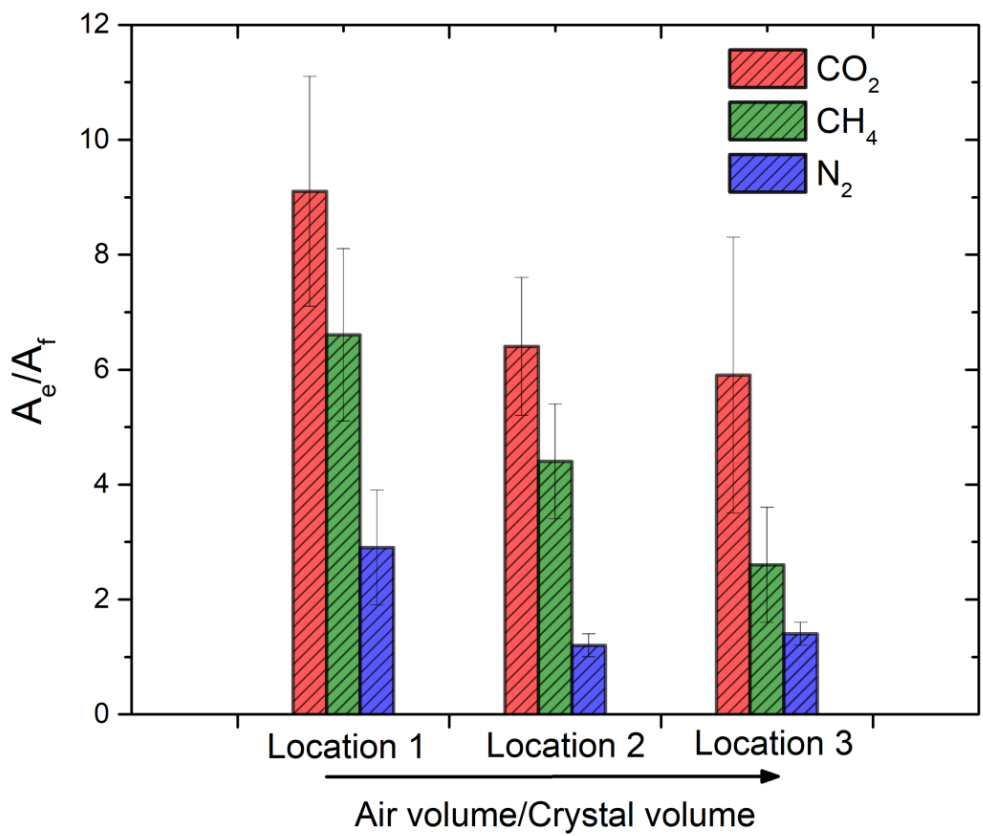

**Figure 4**. Bar graphic showing the relative area of bound gases ($A_e$) normalized by the free gas ($A_f$). The decrease in the magnitude implies a weaker interaction. Each group corresponds to a different focus and crystals, therefore the decrease in the amplitude between groups accounts for different air/ crystal volume measured.

The resulting selectivity values for $CO_2/CH_4$, $CO_2/N_2$, and $CH_4/N_2$, are listed in Table 2 for all three sample locations. While subject to uncertainties due to signal-to-noise ratios and limited spectral resolution, the average selectivities $CO_2/CH_4 = 1.5$, $CO_2/N_2 = 4.5$, and $CH_4/N_2 = 2.5$ clearly indicate competitive binding among the gases, with a preference of Cryptophane-A for $CO_2 > CH_4 > N_2$.

**Table 2.** Selectivity of Cryptophane-A for ($CO_2$/$CH_4$/$N_2$).

| Gases | Location (1) | Location (2) | Location (3) | Average |
|---|---|---|---|---|
| $CO_2$/$CH_4$ | 1.5±0.4 | 1.4±0.4 | 2.2±1.2 | 1.5±0.3 |
| $CO_2$/$N_2$ | 5.5±1.3 | 3.2±1.8 | 4.2±1.8 | 4.5±0.9 |
| $CH_4$/$N_2$ | 3.7±1.0 | 2.3±0.9 | 1.9±0.8 | 2.5±0.5 |

The binding preference correlates with the gases' boiling points ($N_2$: –195°C < $CH_4$: –161°C < $CO_2$: –78°C) and their corresponding polarizabilities ($N_2$: 1.7 Å³, $CH_4$: 2.6 Å³, $CO_2$: 2.9 Å³).[20] Higher polarizability generally enhances van der Waals interactions and the ability to induce dipoles in the surrounding environment, such as within the Cryptophane-A cavity. Similar effects have been observed in earlier studies comparing binding of $CHCl_3$ (8.129 Å³) and $CHF_2Cl$ (4.440 Å³).[20] The binding preference also increases according to molecular volume ($N_2$: 26 Å³, $CH_4$: 28 Å³, $CO_2$: 32 Å³), although the differences between these three molecules are small, and all three molecules have a much smaller volume than other guests known for strong binding with cryptophane-A like chloroform (72 Å³) or xenon (42 Å³). While polarizability and size are often correlated, we expect molecular size to play a lesser role in this system.

In addition to integrated peak intensities, spectral broadening and red shifts of the Raman peaks offer further insight into host–guest interactions. The trapped methane exhibits particularly strong broadening compared to its free-state spectrum, which is consistent with confinement effects such as collisional broadening, as reported for methane clathrates[21]. Although broadening is often related to the ratio of cavity size to molecular size, other factors — including molecular symmetry and relaxation/dephasing mechanisms — also contribute. For instance, despite its larger size, encapsulated $CO_2$ shows less broadening than methane, possibly due to its linear geometry and different vibrational relaxation dynamics. Trapped nitrogen shows the least broadening, consistent with its smaller size. Interestingly, the free nitrogen peak appears broader than expected, possibly due to a fraction of nitrogen bound at intermediate positions in Cryptophane-A, i.e. interacting but not fully encapsulated. Such interaction may result in slightly red-shifted Raman signal relative to the free-state, but unresolved in the spectra. If present, this could mean nitrogen's interaction with Cryptophane-A is somewhat underestimated.

Although Raman studies on cryptophane-A complexes with $N_2$, $CH_4$, and $CO_2$ have not yet been reported, the observed red shift for encapsulated methane (26.8 cm$^{-1}$) is in a good agreement with previously measured values for chloroform–Cryptophane-A complexes (28 cm$^{-1}$),[22] and exceeds that reported for methane clathrates at low temperatures (21.2 cm$^{-1}$).[21] This suggests a relatively strong host-guest interaction and a related weakening of the C–H bond. Similarly $CO_2$ hydrates synthesized at high pressures and low temperatures show a Raman red shift of 8.5 cm$^{-1}$ , significantly smaller that cryptophane-A / $CO_2$ complexes.[23] However, binding constants and vibrational red shifts do not always correlate directly, and further validation through computational simulations would be necessary. [24]

It is important to note that simultaneous encapsulation of two guest molecules within a single cage is highly unlikely. Dual occupancy has not been observed for methane in solution and is even less likely for nitrogen due to its lower affinity and consequently shorter residence time within the Cryptophane-A cavity. In the case of $CO_2$, its larger molecular dimension as well reduces the likelihood of co-occupancy within the cage.

**Conclusions**

We employed Raman spectroscopy for the first time as a direct and effective method to demonstrate Cryptophane-A's affinity for methane at room temperature, eliminating the need to extrapolate from low-temperature solution data. Moreover, we showed that the Cryptophane-A cavity is accessible not only to methane but also to nitrogen and carbon dioxide—two gases that have not been previously studied with this host. In all cases, gas molecules encapsulated within the cryptophane cage exhibited redshifted Raman signals relative to their free-gas counterparts, consistent with positive host–guest interactions and van der Waals forces.

In addition, we quantified the relative selectivity of Cryptophane-A for these gases, confirming that, under our experimental conditions, the host exhibits approximately 1.5 times greater affinity for $CO_2$ than for $CH_4$. While methane's affinity is higher than that of nitrogen, which we independently verified using Quartz Crystal Microbalance (QCM) measurements (see Supplementary Information S1) and which aligns with numerous studies on cryptophane-A doped polymers for methane sensing, the relatively modest selectivity factor of 2.5 suggests significant competitive binding between methane and nitrogen. This insight is particularly important, given that nitrogen is frequently used as a carrier gas in sensing applications, yet its interaction with cryptophanes has often been overlooked. As for $CO_2$, similar encapsulation behavior has been observed in metal–organic frameworks (MOFs) only under high-pressure and low-temperature conditions, suggesting that molecular cages[25] such as cryptophanes possess a stronger trapping capability for carbon dioxide under ambient conditions.[26] Binding of $CO_2$ to cryptophane, although indirectly measured for cryptophane 111, [15] has been dismissed in literature. QCM measurement following previously reported claddings for $CO_2$ binding to cryptophane-A can be found in Supplementary Information S1.

Our findings suggest that the cavity size of Cryptophane-A should be treated as an upper bound, enabling molecules with van der Waals volumes below ~95 $Å^3$ to interact with the host.[4] This implies potential interferents such as water, ethanol, nitrogen, carbon dioxide, halomethanes, acetylene, ethane, xenon, argon, nitrogen oxides, ammonia, and others, raising significant implications for sensor design. In this context, Cryptophane-A-based sensors cannot be considered methane-specific, and pairing them with non-selective transduction methods (e.g., refractive index or mass sensors) is unreliable and unsuitable for environmental monitoring.

Therefore, analytical specificity must be ensured using techniques such as IR, Raman, or mass spectrometry, particularly when analyzing unknown mixtures or when the relative affinities of the target gases are not well characterized. Nevertheless, cryptophanes, and Cryptophane-A in particular, remain valuable as preconcentration agents in combination with these techniques, offering potential for substantial improvements in detection limits.

Overall, Raman spectroscopy proved to be a highly specific and versatile tool for investigating gas encapsulation and host–guest interactions. The method can be extended to other environmentally relevant gases, such as hydrogen and non-protic species that are challenging to study using NMR. Beyond Cryptophane-A, this approach is well-suited to examining other cage-like or porous host systems, including cyclodextrins, calixarenes, pillararenes, cucurbiturils, and materials such as MOFs (metal–organic frameworks), COFs (covalent organic frameworks), HOFs (hydrogen-bonded organic frameworks), and PIMs (polymers of intrinsic microporosity).

**Corresponding Author**

* sebastian.alberti@uit.no

**Author Contributions**

S.A. Designed and performed the experiments and carried out the data analysis. T.B. Synthetized Cryptophane-A. J.J supervised the project and secured funding. The manuscript was written through contributions of all authors. All authors have given approval to the final version of the manuscript.

**Funding Sources**

This work was supported by the European Research Council (grant no. 758973) and Tromsø Research Foundation (project ID 17_SG_JJ).

ACKNOWLEDGMENT

(Word Style "TD_Acknowledgments"). Generally the last paragraph of the paper is the place to acknowledge people, organizations, and financing (you may state grant numbers and sponsors here). Follow the journal's guidelines on what to include in the Acknowledgments section.

REFERENCES

(1)     Garel, L.; Dutasta, J.; Collet, A. Complexation of Methane and Chlorofluorocarbons by Cryptophane-A in Organic Solution. *Angew. Chemie Int. Ed. English* **1993**, *32* (8), 1169–1171. https://doi.org/10.1002/anie.199311691.

(2)     Garel, L.; Lozach, B.; Dutasta, J. P.; Collet, A. Remarkable Effect of the Receptor Size in the Binding of Acetylcholine and Related Ammonium Ions to Water-Soluble Cryptophanes. *J. Am. Chem. Soc.* **1993**, *115* (24), 11652–11653. https://doi.org/10.1021/ja00077a096.

(3)     Brotin, T.; Berthault, P.; Pitrat, D.; Mulatier, J.-C. Selective Capture of Thallium and Cesium by a Cryptophane Soluble at Neutral PH. *J. Org. Chem.* **2020**, *85* (15), 9622–9630. https://doi.org/10.1021/acs.joc.0c00950.

(4)     Huber, G.; Beguin, L.; Desvaux, H.; Brotin, T.; Fogarty, H. A.; Dutasta, J. P.; Berthault, P.

Cryptophane-Xenon Complexes in Organic Solvents Observed through NMR Spectroscopy. *J. Phys. Chem. A* **2008**, *112* (45), 11363–11372. https://doi.org/10.1021/jp807425t.

(5)     Yang, J.; Zhou, L.; Huang, J.; Tao, C.; Li, X.; Chen, W. Sensitivity Enhancing of Transition Mode Long-Period Fiber Grating as Methane Sensor Using High Refractive Index Polycarbonate/Cryptophane A Overlay Deposition. *Sensors Actuators, B Chem.* **2015**, *207* (Part A), 477–480. https://doi.org/10.1016/j.snb.2014.10.013.

(6)     Zhang, Y.; Zhao, Y.; Wang, Q. Measurement of Methane Concentration with Cryptophane E Infiltrated Photonic Crystal Microcavity. *Sensors Actuators B Chem.* **2015**, *209*, 431–437. https://doi.org/10.1016/j.snb.2014.12.002.

(7)     Dullo, F. T.; Lindecrantz, S.; Jágerská, J.; Hansen, J. H.; Engqvist, M.; Solbø, S. A.; Hellesø, O. G. Sensitive On-Chip Methane Detection with a Cryptophane-A Cladded Mach-Zehnder Interferometer. *Opt. Express* **2015**, *23* (24), 31564. https://doi.org/10.1364/OE.23.031564.

(8)     Estelmann, A.; Prien, R.; Marz, W.; Elbing, M.; Harz, P.; Rehder, G. An SPR-Based In Situ Methane Sensor for the Aqueous and Gas Phase. *Anal. Chem.* **2024**. https://doi.org/10.1021/acs.analchem.4c02875.

(9)     Sun, P.; Jiang, Y.; Xie, G.; Du, X.; Hu, J. A Room Temperature Supramolecular-Based Quartz Crystal Microbalance (QCM) Methane Gas Sensor. *Sensors Actuators, B Chem.* **2009**, *141* (1), 104–108. https://doi.org/10.1016/j.snb.2009.06.012.

(10)    Mough, S. T.; Goeltz, J. C.; Holman, K. T. Isolation and Structure of an "Imploded" Cryptophane. *Angew. Chemie - Int. Ed.* **2004**, *43* (42), 5631–5635.

https://doi.org/10.1002/anie.200460866.

(11)  Collet, A. Cyclotriveratrylenes and Cryptophanes. *Tetrahedron* **1987**, *43* (24), 5725–5759. https://doi.org/10.1016/S0040-4020(01)87780-2.

(12)  Bouchet, A.; Brotin, T.; Linares, M.; Cavagnat, D.; Buffeteau, T. Influence of the Chemical Structure of Water-Soluble Cryptophanes on Their Overall Chiroptical and Binding Properties. *J. Org. Chem.* **2011**, *76* (19), 7816–7825. https://doi.org/10.1021/jo201167w.

(13)  Eggers, D. K.; Fu, S.; Ngo, D. V.; Vuong, E. H.; Brotin, T. Thermodynamic Contribution of Water in Cryptophane Host–Guest Binding Reaction. *J. Phys. Chem. B* **2020**, *124* (30), 6585–6591. https://doi.org/10.1021/acs.jpcb.0c05354.

(14)  Schramm, U.; Roesky, C. E. .; Winter, S.; Rechenbach, T.; Boeker, P.; Schulze Lammers, P.; Weber, E.; Bargon, J. Temperature Dependence of an Ammonia Sensor in Humid Air Based on a Cryptophane-Coated Quartz Microbalance. *Sensors Actuators B Chem.* **1999**, *57* (1–3), 233–237. https://doi.org/10.1016/S0925-4005(99)00084-2.

(15)  Chaffee, K. E.; Fogarty, H. A.; Brotin, T.; Goodson, B. M.; Dutasta, J. P. Encapsulation of Small Gas Molecules by Cryptophane-111 in Organic Solution. 1. Size-and Shape-Selective Complexation of Simple Hydrocarbons. *J. Phys. Chem. A* **2009**, *113* (49), 13675–13684. https://doi.org/10.1021/jp903452k.

(16)  Souteyrand, E.; Nicolas, D.; Martin, J. R.; Chauvet, J. P.; Collet, A.; Dutesta, J. P.; Perez, H.; Armand, F. Behaviour of Cryptophane Molecules in Gas Media. *Int. Conf. Solid-State Sensors Actuators, Eurosensors IX, Proc.* **1995**, *1*, 882–885. https://doi.org/10.1109/sensor.1995.717374.

(17) Little, M. A.; Donkin, J.; Fisher, J.; Halcrow, M. A.; Loder, J.; Hardie, M. J. Angewandte Host – Guest Chemistry Synthesis and Methane-Binding Properties of Disulfide-Linked. **2012**, 764–766. https://doi.org/10.1002/anie.201106512.

(18) Brotin, T.; Devic, T.; Lesage, A.; Emsley, L.; Collet, A. Synthesis of Deuterium-Labeled Cryptophane-A and Investigation of Xe@cryptophane Complexation Dynamics by 1D-EXSY NMR Experiments. *Chem. - A Eur. J.* **2001**, *7* (7), 1561–1573. https://doi.org/10.1002/1521-3765(20010401)7:7<1561::AID-CHEM1561>3.0.CO;2-9.

(19) Petrov, D. V.; Matrosov, I. I.; Zaripov, A. R. Determination of Atmospheric Carbon Dioxide Concentration Using Raman Spectroscopy. *J. Mol. Spectrosc.* **2018**, *348*, 137–141. https://doi.org/10.1016/j.jms.2018.01.001.

(20) Https://Cccbdb.Nist.Gov/Pollistx.Asp?

(21) Li, M.; Li, K.; Yang, L.; Su, Y.; Zhao, J.; Song, Y. Evidence of Guest-Guest Interaction in Clathrates Based on in Situ Raman Spectroscopy and Density Functional Theory. *J. Phys. Chem. Lett.* **2022**, *13* (1), 400–405. https://doi.org/10.1021/acs.jpclett.1c03857.

(22) Cavagnat, D.; Brotin, T.; Bruneel, J. L.; Dutasta, J. P.; Thozet, A.; Perrin, M.; Guillaume, F. Raman Microspectrometry as a New Approach to the Investigation of Molecular Recognition in Solids: Chloroform-Cryptophane Complexes. *J. Phys. Chem. B* **2004**, *108* (18), 5572–5581. https://doi.org/10.1021/jp0375158.

(23) Chen, L.; Lu, H.; Ripmeester, J. A. Raman Spectroscopic Study of CO 2 in Hydrate Cages. *Chem. Eng. Sci.* **2015**, *138*, 706–711. https://doi.org/10.1016/j.ces.2015.09.001.

(24)   Yao, Y.; Nijem, N.; Li, J.; Chabal, Y. J.; Langreth, D. C.; Thonhauser, T. Analyzing the Frequency Shift of Physiadsorbed $CO_2$ in Metal Organic Framework Materials. *Phys. Rev. B - Condens. Matter Mater. Phys.* **2012**, *85* (6), 064302. https://doi.org/10.1103/PhysRevB.85.064302.

(25)   Jin, Y.; Voss, B. A.; Jin, A.; Long, H.; Noble, R. D.; Zhang, W. Highly $CO_2$ -Selective Organic Molecular Cages: What Determines the $CO_2$ Selectivity. *J. Am. Chem. Soc.* **2011**, *133* (17), 6650–6658. https://doi.org/10.1021/ja110846c.

(26)   Hu, Y.; Liu, Z.; Xu, J.; Huang, Y.; Song, Y. Evidence of Pressure Enhanced $CO_2$ Storage in ZIF-8 Probed by FTIR Spectroscopy. *J. Am. Chem. Soc.* **2013**, *135* (25), 9287–9290. https://doi.org/10.1021/ja403635b.